\documentclass[12pt]{amsart}
\usepackage{amsmath,amssymb,hyperref,enumitem,graphicx}
\usepackage[sort&compress,numbers]{natbib}

\hypersetup{colorlinks=true,allcolors=blue}
\usepackage[all]{hypcap}

\usepackage[foot]{amsaddr}
\usepackage[sc]{mathpazo}

\newcommand\ack{\subsection*{Acknowledgment}}

\DeclareMathAlphabet\mathsfbi{T1}{phv}{b}{it}

\numberwithin{equation}{section}

\newcommand\BV{\boldsymbol} 
\newcommand\BM{\mathsfbi} 

\newcommand\dif{\:\!\mathrm{d}}
\newcommand\deriv[2]{\frac{\mathrm{d} #1}{\mathrm{d} #2}}
\newcommand\parderiv[2]{\frac{\partial #1}{\partial #2}}

\newcommand\trace{\mathrm{tr}}

\newcommand\myatop[2]{\genfrac{}{}{0pt}{}{#1}{#2}}

\def \Mach{\textit{Ma}}
\def \Rey {\textit{Re}}
\def \Nu {\mathcal V}

\begin{document}

\author[Rafail V. Abramov]{Rafail V. Abramov}

\address{Department of Mathematics, Statistics and Computer Science,
University of Illinois at Chicago, 851 S. Morgan st., Chicago, IL 60607}

\email{abramov@uic.edu}

\title{Macroscopic turbulent flow via hard sphere potential}

\begin{abstract}
In recent works, we proposed a hypothesis that the turbulence in gases
could be produced by particles interacting via a potential, and
examined the proposed mechanics of turbulence formation in a simple
model of two particles for a variety of different potentials. In this
work, we use the same hypothesis to develop new fluid mechanics
equations which model turbulent gas flow on a macroscopic scale. The
main difference between our approach and the conventional formalism is
that we avoid replacing the potential interaction between particles
with the Boltzmann collision integral. Due to this difference, the
velocity moment closure, which we implement for the shear stress and
heat flux, relies upon the high Reynolds number condition, rather than
the Newton law of viscosity and the Fourier law of heat
conduction. The resulting system of equations of fluid mechanics
differs considerably from the standard Euler and Navier--Stokes
equations. A numerical simulation of our system shows that a laminar
Bernoulli jet of an argon-like hard sphere gas in a straight pipe
rapidly becomes a turbulent flow. The time-averaged Fourier spectra of
the kinetic energy of this flow exhibit Kolmogorov's negative
five-thirds power decay rate.
\end{abstract}

\maketitle

\section{Introduction}

\citet{Rey83} demonstrated that an initially laminar flow in a long
straight pipe develops turbulent motions whenever the high Reynolds
number condition is satisfied. More than half a century later,
\citet{Kol41a,Kol41b,Kol41c} and \citet{Obu41,Obu49,Obu62} observed
that the time-averaged Fourier spectra of the kinetic energy of a
turbulent flow possess a universal decay structure, corresponding to
the negative five-thirds power of the Fourier wavenumber. To this day,
the mechanism of turbulence formation in a laminar flow, as well as
the origins of the power scaling of its kinetic energy spectrum,
remain unknown.

In our recent work \citep{Abr20}, we proposed a hypothesis that the
turbulence in gases could be produced by their particles interacting
via a potential. Later, in \citep{Abr21}, we studied the ability of
the potential interaction in a pair of particles to form turbulent
structures with power decay spectra from an initially laminar shear
flow. The key difference between the conventional fluid mechanics
equations and our system was the closure for the velocity moment
hierarchy. While the Newton law of viscosity and the Fourier law of
heat conduction are typically used to truncate the velocity moment
hierarchy emerging from the Boltzmann equation
\citep{Bol,CerIllPul,ChaCow,HirCurBir} (which subsequently leads to
the Euler \citep{Bat} and Navier--Stokes \citep{Gols} equations), in
our potential-driven system we used the high Reynolds number condition
to obtain the closures for the shear stress and heat flux.

In \citep{Abr21}, we examined the dynamics for the following common
interaction potentials: electrostatic, gravitational, Thomas--Fermi
\citep{Tho,Fer}, and Lennard-Jones \citep{Len}. For each potential, we
found via numerical simulations that the time-averaged kinetic energy
of the flow decays as the negative five-thirds power of its Fourier
wavenumber, thus corresponding to the Kolmogorov turbulence
spectrum. Out of curiosity, in \citep{Abr21} we also simulated the
dynamics with a generic large scale potential, which did not
correspond to any known type of interaction, but rather
``impersonated'' a generic mean field forcing, which typically arises
in a Vlasov-type equation \citep{Vla}.  To our surprise, we found
that, even for a large scale mean field potential, the Fourier
transform of the kinetic energy of the flow had the Kolmogorov decay
rate.

The present work complements \citep{Abr21}, in the sense that here we
focus entirely on the macroscopic dynamics of a single particle in a
multiparticle system under the averaged forcing from all other
particles. This is similar to the conventional kinetic theory
approach, with the exception that we do not replace the potential
interaction between particles with a Boltzmann collision integral, and
instead leave it as is. Subsequently, we truncate the velocity moment
hierarchy using the high Reynolds number condition, just as we did in
\citep{Abr21}. The result is the system of fluid dynamics equations
for the density, velocity, temperature and pressure, which is markedly
different from the conventional Euler equations. The first difference
is that the momentum equation now includes the mean field forcing,
which is absent from the Euler equations. The second difference is
that our system preserves the pressure along the streamlines, while
the Euler equations preserve the entropy. We subsequently demonstrate,
via a numerical simulations that, in our system, a laminar Bernoulli
jet develops into a turbulent flow which has the Kolmogorov energy
spectrum.

The work is organized as follows. In Section \ref{sec:multiparticle},
we consider a system of many particles which interact via a potential.
For this system, we compute the Liouville equation, the Gibbs
equilibrium state of the system, and show that a generic solution
preserves the family of the R\'enyi metrics \citep{Ren}, including the
Kullback--Leibler divergence \citep{KulLei}, between itself and the
equilibrium state.

In Section \ref{sec:closure} we carry out the
Bogoliubov--Born--Green--Kirkwood--Yvon (BBGKY) formalism
\citep{Bog,BorGre,Kir} to obtain the transport equation for the
distribution density of a single particle. We then close this
transport equation by expressing the two-particle marginal density via
the single-particle density in the same manner as in our recent work
\citep{Abr17}. The result is the Vlasov-type equation with a large
scale mean field forcing. In the case of a short range interaction
potential, we obtain a simplified expression for the mean field
forcing, and, in particular, an explicitly computable formula for the
mean field forcing derived from a hard sphere collision interaction.

In Section \ref{sec:turbulent_equations}, we formulate a hierarchy of
the transport equations for the velocity moments of the resulting
Vlasov-type equation. Due to the fact that the potential forcing
replaces the usual Boltzmann collision integral, a different closure
must be used to truncate the moment hierarchy. Here, we use the same
closure for the shear stress and heat flux, based on the high Reynolds
number condition, as we introduced in our recent work \citep{Abr21}.
This closure leads to the system of transport equations for the
density, momentum, and pressure (or, equivalently, the inverse
temperature).

In Section \ref{sec:numerics} we show the results of a numerical
simulation of the new transport equations in a realistic setting.
Namely, we simulate the flow of argon in a straight pipe. The initial
condition for the flow is a laminar Bernoulli jet, which is a steady
state in the conventional Euler equations. However, in our model, this
laminar Bernoulli jet quickly develops into a turbulent flow, which
behaves visually similarly to turbulent flows observed in nature. To
verify that the behavior of our model is indeed related to turbulent
flows, we compute the time averages of the Fourier transform of the
kinetic energy of the turbulent flow. We observe that these averages
indeed decay with the rate of negative five-thirds power of the
Fourier wavenumber, which is a famous Kolmogorov spectrum. Section
\ref{sec:summary} summarizes the results of this work.

\section{Multiparticle dynamics}
\label{sec:multiparticle}

Here, we consider a dynamical system which consists of $N$ identical
particles, interacting via a potential $\phi(r)$. Denoting the
coordinate and velocity of the $i$-th particle via $\BV x_i$ and $\BV
v_i$, respectively, we have the following system of equations of
motion:
\begin{equation}
\label{eq:dyn_sys_N}
\deriv{\BV x_i}t=\BV v_i,\qquad\deriv{\BV v_i}t=-\sum_{\myatop{j=1}{
    j\neq i}}^N\parderiv{}{\BV x_i}\phi(\|\BV x_i-\BV x_j\|).
\end{equation}
Observe that adding an arbitrary constant to $\phi(r)$ leaves the
equations in \eqref{eq:dyn_sys_N} unaffected. In what follows, for
convenience we assume that this additive constant is such that
$\phi(r)\to 0$ as $r\to\infty$. For the purposes of this work, we also
assume that the potential $\phi(r)$ is repulsive, that is, $\phi(r)\gg
1$ as $r\to 0$. We do not, however, assume that $\phi(r)$ necessarily
becomes infinite at zero, to avoid technical issues associated with
this singularity.

The total momentum and energy of all particles are preserved by the
dynamics:
\begin{equation}
\sum_{i=1}^N\BV v_i=\text{const},\qquad\sum_{i=1}^N\frac{\|\BV v\|^2}
2+\sum_{i=1}^{N-1}\sum_{j=i+1}^N\phi(\|\BV x_i-\BV x_j\|)=
\text{const}.
\end{equation}
Observe that, for a given value of the momentum, it is always possible
to choose the inertial reference frame in which the momentum becomes
zero; thus, without much loss of generality, we will further assume
that the total momentum of the system is zero.

Following our work \citep{Abr20}, we concatenate the coordinates as
$\BV X=(\BV x_1,\ldots,\BV x_N)$, and velocities as $\BV V=(\BV
v_1,\ldots,\BV v_N)$. In these notations, we can write
\begin{equation}
\deriv{\BV X}t=\BV V,\qquad\deriv{\BV V}t=-\parderiv\Phi{\BV X},\qquad
\Phi(\BV X)=\sum_{i=1}^{N-1}\sum_{j=i+1}^N\phi(\|\BV x_i-\BV x_j\|).
\end{equation}
In the variables $\BV X$ and $\BV V$, the conservation of the energy
can be expressed via
\begin{equation}
\|\BV V\|^2+2\Phi(\BV X)=\text{const}.
\end{equation}
Let $F(t,\BV X,\BV V)$ be the density of states of the dynamical
system above. Then, the Liouville equation for $F$ is given via
\begin{equation}
\label{eq:liouville_N}
\parderiv Ft+\BV V\cdot\parderiv F{\BV X}=\parderiv\Phi{\BV X}\cdot
\parderiv F{\BV V}.
\end{equation}
Any suitable $F_0$ of the form
\begin{equation}
F_0(\BV X,\BV V)=F_0\big(\|\BV V\|^2+2\Phi(\BV X)\big)
\end{equation}
is a steady state for \eqref{eq:liouville_N}. Among those, the
canonical Gibbs state is
\begin{equation}
\label{eq:F_G_N}
F_G(\BV X,\BV V)=\frac
1{(2\pi\theta_0)^{3N/2}Z_N}\exp\left(-\frac{\|\BV V\|^2+2\Phi(\BV X)}{
2\theta_0}\right),\qquad Z_N=\int e^{-\Phi(\BV X)/\theta_0}\dif\BV X,
\end{equation}
where $\theta_0$ is the kinetic temperature. As shown by
\citet{Abr21}, \eqref{eq:liouville_N} preserves the family of general
R\'enyi divergences \citep{Ren}; indeed, for $\psi_1(F)$ and
$\psi_2(F)$ we have
\begin{multline}
\parderiv{}t\int\psi_1(F)\psi_2(F_0)\dif\BV X\dif\BV V=\int\psi_2(
F_0)\parderiv{\psi_1(F)}t\dif\BV X\dif\BV V=\int\psi_2(F_0)
\bigg(\parderiv\Phi{\BV X}\cdot\parderiv{\psi_1(F)}{\BV V}-\\-\BV
V\cdot\parderiv{\psi_1(F)}{\BV X}\bigg) \dif\BV X\dif\BV
V=\int\psi_1(F)\bigg(\BV V\cdot\parderiv{\psi_2(F_0)}{\BV
  X}-\parderiv\Phi{\BV X}\cdot \parderiv{\psi_2(F_0)}{\BV
  V}\bigg)\dif\BV X\dif\BV V=0,
\end{multline}
and, if $\psi_1(x)=x^\alpha$, $\psi_2(x)=x^{1-\alpha}$ for some
$\alpha>0$, the following R\'enyi divergence is preserved in time:
\begin{equation}
D_\alpha(F,F_0)=\frac 1{\alpha-1}\ln\int F^\alpha F_0^{1-\alpha}
\dif\BV X\dif\BV V.
\end{equation}
The Kullback--Leibler divergence \citep{KulLei} is a special case of
the R\'enyi divergence for $\alpha=1$. As we remarked in
\citep{Abr21}, the conservation of the R\'enyi metrics plays a similar
role to Boltzmann's $H$-theorem -- namely, if the initial condition
of~\eqref{eq:liouville_N} is close, in the sense of any R\'enyi
metric, to \eqref{eq:F_G_N}, then the solution will also remain a
nearby state.

\subsection{Two-particle marginal density of the Gibbs state}

The marginal density of the Gibbs state for the particles \#1 and \#2
is given via
\begin{multline}
F^{(2)}_G(\BV x_1,\BV v_1,\BV x_2,\BV v_2)=\int F_G(\BV X,\BV
V)\dif\BV x_3\dif\BV v_3\ldots\dif\BV x_N\dif\BV v_N=\frac
1{(2\pi\theta_0)^3}e^{-(\|\BV v_1\|^2+\|\BV v_2\|^2)/2\theta_0}\\e^{
  -\phi(\|\BV x_2-\BV x_1\|)/\theta_0}Y^{(2)}_N(\|\BV x_2-\BV x_1\|)
=f_G(\BV v_1)f_G(\BV v_2)e^{-\phi(\|\BV x_2-\BV x_1\|)/\theta_0}V^2
Y^{(2)}_N(\|\BV x_2-\BV x_1\|),
\end{multline}
where $V$ is the volume of the coordinate domain, $f_G(\BV v)$ is a
single-particle marginal Gibbs density,
\begin{equation}
f_G(\BV v)=\frac 1{(2\pi\theta_0)^{3/2}V}e^{-\|\BV v\|^2/2\theta_0},
\end{equation}
and $Y^{(2)}_N(r)$ is a multiple of the two-particle cavity
distribution function \citep{Bou06}:
\begin{multline}
\label{eq:cavity_DF}
Y^{(2)}_N(\|\BV x_2-\BV x_1\|)=\frac 1{Z_N}\int\bigg(\prod_{i=3}^N
e^{-(\phi(\|\BV x_i-\BV x_1\|)+\phi(\|\BV x_i-\BV x_2\|))/\theta_0}
\\\prod_{j=i+1}^N e^{-\phi(\|\BV x_i-\BV x_j\|)/\theta_0} \bigg)
\dif\BV x_3\ldots\dif\BV x_N.
\end{multline}
The gas is {\em dilute} if the measure of the set where the integrand
of \eqref{eq:cavity_DF} is not unity is negligible in comparison with
the measure of the set over which the integration occurs. In such a
case, $V^2Y^{(2)}_N=1$, which leads to the simplified expression for
the two-particle marginal density:
\begin{equation}
\label{eq:F_G^2}
F^{(2)}_G(\BV x_1,\BV v_1,\BV x_2,\BV v_2)=f_G(\BV v_1)f_G(\BV v_2)
e^{-\phi(\|\BV x_2-\BV x_1\|)/\theta_0}.
\end{equation}

\section{The closure for a single particle (Vlasov equation)}
\label{sec:closure}

Here, we isolate a single particle (say, \#1), and denote $\BV x=\BV
x_1$, $\BV v=\BV v_1$. We then examine the transport of its marginal
distribution $f(t,\BV x,\BV v)$, given via
\begin{equation}
f(t,\BV x,\BV v)=\int F(t,\BV X,\BV V)\dif\BV x_2\dif\BV v_2\ldots
\dif\BV x_N\dif\BV v_N.
\end{equation}
Integrating the Liouville equation in \eqref{eq:liouville_N} over all
particles but the first one, in the absence of boundary effects we
arrive at
\begin{equation}
\parderiv ft+\BV v\cdot\parderiv f{\BV x}=\sum_{i=2}^N\int
\parderiv{}{\BV x}\phi(\|\BV x-\BV y\|) \cdot \parderiv{}{\BV
  v}F^{(2)}_{1,i}(\BV x,\BV v,\BV y,\BV w)\dif\BV y\dif\BV w,
\end{equation}
where $F^{(2)}_{1,i}$ is the two-particle marginal density for
particles $1$ and $i$. The equation above is a part of the BBGKY
hierarchy \citep{Bog,BorGre,Kir}, which needs to be {\em closed}, that
is, $F^{(2)}_{1,i}$ needs to be approximated via $f$. In order to do
this, we assume that all pairs of particles are statistically
identical, and thus their joint marginal distributions are equal. This
leads to
\begin{equation}
\parderiv ft+\BV v\cdot\parderiv f{\BV x}=(N-1)\int\parderiv{}{\BV
  x}\phi(\|\BV x-\BV y\|) \cdot \parderiv{}{\BV v}F^{(2)}(\BV x,\BV
v,\BV y,\BV w)\dif\BV y\dif\BV w,
\end{equation}
where $F^{(2)}$ is the two-particle marginal density for any pair of
particles. Here, the simplest closure would, apparently, be the direct
factorization of $F^{(2)}$ into the product of $f$'s, however, such
closure would not be compatible with the form of the steady state
marginal distribution in \eqref{eq:F_G^2}. Instead, we use the same
type of closure as we previously did in~\citep{Abr17}. Namely, we take
the exact expression for the two-particle equilibrium state of a
dilute gas in \eqref{eq:F_G^2}, and replace, first, the single
particle equilibrium marginals $f_G$ with $f$, and, second, the
equilibrium kinetic temperature $\theta_0$ with that of $f$, computed
at the midpoint between $\BV x$ and $\BV y$:
\begin{equation}
F^{(2)}(\BV x,\BV v,\BV y,\BV w)=f(\BV x,\BV v)f(\BV y,\BV w)\exp
\bigg(-\frac{\phi(\|\BV x-\BV y\|)}{\theta(\frac{\BV x+\BV y}2)}
\bigg).
\end{equation}
This leads to the closed transport equation for $f$:
\begin{subequations}
\begin{equation}
\parderiv ft+\BV v\cdot\parderiv f{\BV x}=\BV g\cdot\parderiv f{\BV
  v},
\end{equation}
\begin{equation}
\BV g=(N-1)\int\exp \bigg(-\frac{\phi(\|\BV x-\BV y\|)}{\theta(\frac{
    \BV x+\BV y}2)} \bigg)\parderiv{}{\BV x}\phi(\|\BV x-\BV y\|)f(\BV
y,\BV w)\dif\BV y\dif \BV w.
\end{equation}
\end{subequations}
This is an equation of the Vlasov type \citep{Vla}, with the mean
field forcing provided by $\BV g$. Following the standard approach
\citep{Abr17}, we renormalize $f$ so that it becomes a mass density,
$f\to Nmf$, where $m$ is the mass of a particle. The transport
equation for $f$ remains the same, however, $\BV g$ becomes
renormalized as follows:
\begin{equation}
\BV g=\frac{N-1}{Nm}\int\exp \bigg(-\frac{\phi(\|\BV x-\BV y\|)}{
  \theta(\frac{ \BV x+\BV y}2)} \bigg) \parderiv{}{\BV x}\phi(\|\BV
x-\BV y\|)f(\BV y,\BV w)\dif\BV y\dif\BV w.
\end{equation}
Next, we assume that $N$ is large enough so that $(N-1)/N\approx 1$,
which yields
\begin{equation}
\BV g=\frac 1m\int\exp \bigg(-\frac{\phi(\|\BV x-\BV y\|)}{ \theta(
  \frac{\BV x+\BV y}2)}\bigg)\parderiv{}{\BV x}\phi(\|\BV x-\BV y\|)
\rho(\BV y)\dif\BV y,\qquad\rho(\BV x)=\int f(\BV x,\BV v)\dif\BV v.
\end{equation}

\subsection{Short range potential}

If $\phi$ is a short range potential, the integral above can be
simplified considerably. Let us change the dummy variable of
integration $\BV y\to\BV x-\BV y$:
\begin{multline}
\int\exp \bigg(-\frac{\phi(\|\BV x-\BV y\|)}{ \theta( \frac{\BV x+\BV
    y}2)}\bigg)\parderiv{}{\BV x}\phi(\|\BV x-\BV y\|) \rho(\BV
y)\dif\BV y=\int\exp \bigg(-\frac{\phi(\|\BV x-\BV y\|)}{ \theta(
  \frac{\BV x+\BV y}2)}\bigg)\\\phi'(\|\BV x-\BV y\|)\frac{\BV x-\BV
  y}{\|\BV x-\BV y\|} \rho(\BV y)\dif\BV
y=\int\exp\bigg(-\frac{\phi(\|\BV y\|)}{\theta(\BV x-\BV y/2)}\bigg)
\phi'(\|\BV y\|)\frac{\BV y}{\|\BV y\|} \rho(\BV x-\BV y)\dif\BV y.
\end{multline}
Next, let us observe that
\begin{multline}
\exp\bigg(-\frac{\phi(\|\BV y\|)}{\theta(\BV x-\BV y/2)}\bigg)\phi'(
\|\BV y\|)\frac{\BV y }{\|\BV y\|}=\exp\bigg(-\frac{\phi(\|\BV y\|)}{
  \theta(\BV x-\BV y/2)}\bigg)\parderiv{\phi(\|\BV y\|)}{\BV y}=\\=
\theta(\BV x-\BV y/2)\exp\bigg(-\frac{\phi(\|\BV y\|)}{\theta(\BV x-
  \BV y/2)}\bigg)\bigg[\parderiv{}{\BV y}\bigg(\frac{\phi(\|\BV y\|)
  }{\theta(\BV x-\BV y/2)}\bigg)+\frac 12\parderiv{}{\BV x}\bigg(
  \frac{\phi(\|\BV y\|)}{\theta(\BV x-\BV y/2)}\bigg)\bigg]=\\=
\theta(\BV x-\BV y/2)\bigg(\parderiv{}{\BV y}+\frac 12\parderiv{}{\BV
  x}\bigg) \bigg[1-\exp\bigg(-\frac{\phi(\|\BV y\|)}{\theta(\BV x-\BV
    y/2)}\bigg)\bigg],
\end{multline}
where in the last identity we add 1 to the quantity under the
differentiation, so that the expression in square brackets approaches
zero as $\|\BV y\|\to\infty$. This leads, via integration by parts, to
\begin{multline}
\int\exp\bigg(-\frac{\phi(\|\BV y\|)}{\theta(\BV x-\BV y/2)}\bigg)
\phi'(\|\BV y\|)\frac{ \BV y}{\|\BV y\|}\rho(\BV x-\BV y)\dif\BV y=
\int\rho(\BV x-\BV y) \theta(\BV x-\BV y/2)\\\bigg(\parderiv{}{\BV y}
+\frac 12\parderiv{}{\BV x}\bigg)\bigg[1-\exp\bigg(-\frac{\phi(\|\BV y
    \|)}{\theta(\BV x-\BV y /2)}\bigg)\bigg]\dif\BV y=\frac 12\int
\bigg[1-\exp\bigg(-\frac{\phi(\|\BV y\|)}{\theta(\BV x-\BV y/2)}
  \bigg)\bigg]\\\theta(\BV x-\BV y/2)\parderiv{}{\BV x}\rho(\BV x-\BV
y)\dif\BV y+\frac 12\parderiv{}{\BV x}\int\bigg[1-\exp\bigg(-\frac{
    \phi(\|\BV y\|}{\theta(\BV x-\BV y/2)}\bigg)\bigg]\rho(\BV x-\BV
y) \theta(\BV x-\BV y/2)\dif\BV y=\\=\frac 12\int\frac 1{\rho(\BV
  x-\BV y)}\parderiv{}{\BV x}\bigg\{\bigg[ 1-\exp\bigg(-\frac{
    \phi(\|\BV y\|)}{\theta(\BV x-\BV y/2)}\bigg)\bigg]\rho^2(\BV
x-\BV y)\theta(\BV x-\BV y/2)\bigg\}\dif\BV y.
\end{multline}
Observe that $1-e^{-\phi/\theta}\sim\phi/\theta$ as $\phi\to 0$, yet
is bounded above by 1 even if $\phi$ is arbitrarily large at zero.
Thus, if the effective range of $\phi$ is short enough so that $\rho$,
$\theta$ and their spatial derivatives can be treated as constants
within it, we can ignore their dependence on the dummy variable of
integration $\BV y$:
\begin{multline}
\int\frac 1{\rho(\BV x-\BV y)}\parderiv{}{\BV x}\bigg\{\bigg[1-\exp
  \bigg(-\frac{ \phi(\|\BV y\|)}{\theta(\BV x-\BV y/2)} \bigg)\bigg]
\rho^2(\BV x-\BV y)\theta(\BV x-\BV y/2)\bigg\}\dif\BV y=\\=\frac 1{
  \rho(\BV x)}\parderiv{}{\BV x}\bigg\{\rho^2(\BV x)\theta(\BV x)
\int\bigg[1-\exp\bigg(-\frac{\phi(\|\BV y\|)}{\theta(\BV x)}
  \bigg)\bigg]\dif\BV y\bigg\}.
\end{multline}
This leads to
\begin{equation}
\BV g=\frac 1{2m\rho(\BV x)}\parderiv{}{\BV x}\bigg\{\rho^2(\BV x)
\theta(\BV x)\int\bigg[1-\exp\bigg(-\frac{\phi(\|\BV y\|)}{\theta(\BV
    x)}\bigg)\bigg]\dif\BV y\bigg\}.
\end{equation}
Denoting, for further convenience, the mean field potential $\bar\phi$
via
\begin{equation}
\label{eq:mean_field}
\bar\phi(\BV x)=\frac{\rho^2(\BV x)\theta(\BV x)}{2m}\int\bigg[1-\exp
  \bigg(-\frac{\phi(\|\BV y\|)}{\theta(\BV x)}\bigg)\bigg]\dif\BV y,
\end{equation}
we obtain the following Vlasov-type transport equation for a short
range interaction potential $\phi$:
\begin{equation}
\label{eq:vlasov}
\parderiv ft+\BV v\cdot\parderiv f{\BV x}=\frac 1\rho
\parderiv{\bar\phi}{\BV x}\cdot\parderiv f{\BV v}.
\end{equation}
Here, observe that, generally, the mean field potential $\bar\phi$ is
time-dependent, since $\rho$ and $\theta$ are time-dependent.

\subsection{Special case: the mean field potential for a hard sphere
particle}

If $\phi$ is the hard-sphere potential (that is, it is zero outside
the radius of the sphere, and arbitrarily large inside, with a rapid,
yet smooth, transition), then the expression for the mean field
potential in~\eqref{eq:mean_field} can be simplified even
further. Indeed, observe that, for a hard sphere potential, the
integral over $\dif\BV y$ in \eqref{eq:mean_field} is simply the
volume of the hard sphere $V_{sp}$ irrespectively of the value of
$\theta(\BV x)$:
\begin{equation}
\int\bigg[1-\exp\bigg(-\frac{\phi_{HS}(\|\BV y\|)}{\theta(\BV x)}
  \bigg)\bigg]\dif\BV y=V_{sp}.
\end{equation}
Substituting the above expression into \eqref{eq:mean_field}, and
denoting the mass density of such a spherical particle via
$\rho_{sp}=m/V_{sp}$, we obtain the following explicit formula of the
mean field potential for a hard sphere particle of the mass density
$\rho_{sp}$:
\begin{equation}
\label{eq:mean_field_HS}
\bar\phi_{HS}(\BV x)=\frac{\rho^2(\BV x)\theta(\BV x)}{2\rho_{sp}}.
\end{equation}

\section{The transport equations for a turbulent flow}
\label{sec:turbulent_equations}

Just as done in the course of the conventional formalism in fluid
mechanics, here we convert \eqref{eq:vlasov} into a hierarchy of the
transport equations for the velocity moments, which is subsequently
truncated at a suitable point.  However, unlike the conventional fluid
mechanics, here we will use a different closure to truncate the
hierarchy, which is based on the high Reynolds number condition.

As usual, we denote the velocity average $\langle a\rangle$ of a
function $a(\BV v)$ via
\begin{equation}
\langle a\rangle(t,\BV x)=\int a(\BV v)f(t,\BV x,\BV v)\dif\BV v.
\end{equation}
The velocity moments are the averages of powers of $\BV v$, that is
$\langle\BV v^k\rangle$, where $\BV v^k$ is the $k$-fold outer product
(which is a tensor of rank $k$). The first two velocity moments are
the mass density and the momentum, respectively:
\begin{equation}
\rho=\langle 1\rangle,\qquad\rho\BV u=\langle\BV v\rangle.
\end{equation}
The transport equation for any velocity moment is obtained by
multiplying \eqref{eq:vlasov} by an appropriate power of $\BV v$,
integrating over $\dif\BV v$, and using the integration by parts for
the term with $\bar\phi$. In particular, for the density and momentum
we have, respectively,
\begin{equation}
\label{eq:mass_momentum}
\parderiv\rho t+\nabla\cdot(\rho\BV u)=0,\qquad\parderiv{(\rho\BV u)}t
+\nabla\cdot\langle\BV v^2\rangle+\nabla\bar\phi=\BV 0,
\end{equation}
where $\nabla$ now denotes the $\BV x$-differentiation operator (as
the $\BV v$-derivatives are no longer present). Observe that, due to
the presence of $\BV v$ in the advection term of the Vlasov equation
\eqref{eq:vlasov}, each moment equation is coupled to a higher-order
velocity moment (i.e., the equation for $\rho$ is coupled to $\rho\BV
u$, the equation for $\rho\BV u$ is coupled to $\langle\BV
v^2\rangle$, and so forth), thus forming a hierarchy of the velocity
moment transport equations. In order to truncate this hierarchy, a
closure is needed.

To obtain a suitable closure, let us write the transport equation for
the quadratic moment:
\begin{equation}
\label{eq:energy}
\parderiv{\langle\BV v^2\rangle}t+\nabla\cdot\langle\BV v^3\rangle+
\nabla\bar\phi\BV u^T+\BV u\nabla\bar\phi^T=\BM 0,
\end{equation}
where, as above, the terms with the potential forcing were integrated
by parts. Next, we decompose the quadratic and cubic moments as
follows: we introduce the temperature tensor $\BM T$, and express
\begin{subequations}
\begin{equation}
\rho\BM T=\langle(\BV v-\BV u)^2\rangle,\qquad\langle\BV v^2\rangle=
\rho(\BV u^2+\BM T),
\end{equation}
\begin{equation}
\langle\BV v^3\rangle=\rho\big(\BV u^3+\BV u\otimes\BM T+(\BV u\otimes
\BM T)^T+(\BV u\otimes\BM T)^{TT}\big)+\langle(\BV v-\BV u)^3\rangle,
\end{equation}
\end{subequations}
where ``$\otimes$'' denotes the outer product, while ``$T$'' and
``$TT$'' denote the two permutations of a symmetric 3-rank tensor. The
kinetic temperature $\theta$ is the normalized trace of $\BM T$, while
the viscous shear stress $\BM S$ is the corresponding deviator:
\begin{equation}
\theta=\frac 13\trace(\BM T),\qquad\BM S=\BM T-\theta\BM I.
\end{equation}
The product $\rho\theta$ is known as the pressure. By construction,
$\BM S$ is traceless. Likewise, we decompose the centered cubic moment
by introducing the heat flux $\BV q$, and the corresponding deviator
$\BM Q$:
\begin{equation}
\rho\BV q=\frac 12\langle\|\BV v-\BV u\|^2(\BV v-\BV u)\rangle,\quad
\rho\BM Q=\langle(\BV v-\BV u)^3\rangle-\frac 25\rho\big(\BV q\otimes
\BM I+(\BV q\otimes\BM I)^T+(\BV q\otimes\BM I)^{TT}\big).
\end{equation}
By construction, the contraction of $\BM Q$ along any pair of its
indices is zero. In the new variables, the transport equation for the
quadratic moment can be replaced by the transport equations for the
pressure and shear stress:
\begin{subequations}
\label{eq:pressure_stress}
\begin{equation}
\parderiv{(\rho\theta)}t+\nabla\cdot(\rho\theta\BV u)+\frac 23\big(
\rho\theta\nabla\cdot\BV u+\rho\BM S:\nabla\BV u+\nabla\cdot(\rho\BV
q)\big)=0,
\end{equation}
\begin{multline}
\label{eq:stress}
\parderiv{(\rho\BM S)}t+\nabla\cdot(\BV u\otimes\BM S)+\rho\bigg(\BM S
\nabla\BV u+\nabla\BV u^T\BM S-\frac 23(\BM S:\nabla\BV u)\BM I \bigg)
+\nabla\cdot(\rho\BM Q)+\\+\frac 25\bigg(\nabla(\rho\BV q)+\nabla
(\rho\BV q)^T-\frac 23\nabla\cdot(\rho\BV q)\BM I\bigg)+\rho\theta
\bigg(\nabla\BV u+\nabla\BV u^T-\frac 23(\nabla\cdot\BV u)\BM I\bigg)
=\BM 0.
\end{multline}
\end{subequations}
The equations above are obtained by subtracting appropriate
combinations of the mass and momentum transport equations in
\eqref{eq:mass_momentum} from the quadratic moment transport equation
in \eqref{eq:energy} (for more details, see \citet{Gra},
\citet{Abr13,Abr17}, or \citet{AbrOtt}).

\subsection{A turbulent closure for the moment transport equations}

As we can see, the moment transport equations in
\eqref{eq:mass_momentum}, \eqref{eq:energy} and
\eqref{eq:pressure_stress} are chain-linked together via the
higher-order moment in the advection term, forming an infinite
hierarchy of equations. For a practical computation, this hierarchy
must be truncated at some point, so that the number of equations to be
solved is finite.

For the conventional velocity moment hierarchy emerging from the
Boltzmann equation, such truncation typically occurs at the equations
for the shear stress $\BM S$ and heat flux $\BV q$. The reason why it
happens is that the time-irreversible collision integral of the
Boltzmann equation creates linear damping in the shear stress and heat
flux equations, which is inversely proportional to the viscosity. For
a small viscosity, this damping becomes strong, and one can replace
the shear stress and heat flux equations with their respective steady
states, obtaining the Newton law of viscosity, and Fourier law of heat
conduction, respectively. This results in the Navier--Stokes
equations, or, if the viscosity is set to zero (thus making the shear
stress and heat flux damping infinitely strong), in the Euler
equations. For rarefied gas flows with a large Knudsen number, the
truncation can be made at a higher order in the velocity moment
hierarchy, resulting in the 13-moment Grad equations \citep{Gra}, or
the regularized 13-moment Grad equations \citep{StruTor}.

Here, however, observe that such collision damping is not present in
the transport equation for the shear stress \eqref{eq:stress}, thus
necessitating a different closure. Thus, we follow our recent work
\citep{Abr21} and use the high Reynolds number condition to truncate
the moment hierarchy. Let $L$ and $U$ denote, respectively, the
characteristic length scale of the flow and its reference speed, such
that the ratio $U/L$ specifies the characteristic time scale.
Additionally, $T$ will specify the reference temperature of the gas,
while $\Nu$ will be the reference viscosity constant. We rescale the
time and space differentiation operators, as well as the velocity,
temperature, the mean field potential, the shear stress, the heat
flux, and its deviator, via
\begin{subequations}
\begin{equation}
\nabla\to\frac 1L\nabla,\qquad\parderiv{}t\to\frac
UL\parderiv{}t,\quad \BV u\to U\BV u,\qquad\theta\to
T\theta,\qquad\bar\phi\to T\bar\phi,
\end{equation}
\begin{equation}
\BM S\to\frac{\Nu U}L\BM S,\qquad\BV q\to TU\BV q,\qquad\BM Q\to TU\BM Q.
\end{equation}
\end{subequations}
In the rescaled variables, the mass, momentum, pressure and stress
transport equations in \eqref{eq:mass_momentum} and
\eqref{eq:pressure_stress} become
\begin{subequations}
\begin{equation}
\parderiv\rho t+\nabla\cdot(\rho\BV u)=0,\qquad\parderiv{(\rho\BV
  u)}t+\nabla\cdot(\rho\BV u^2)+\frac 1{\Mach^2} \nabla(
\rho\theta+\bar\phi)+\frac 1\Rey\nabla\cdot(\rho\BM S)=\BV 0,
\end{equation}
\begin{equation}
\label{eq:pressure}
\parderiv{(\rho\theta)}t+\nabla\cdot(\rho\theta\BV u)+\frac 23\bigg(
\rho\theta\nabla\cdot\BV u+\frac{\Mach^2}\Rey\rho\BM S:\nabla\BV u+
\nabla\cdot(\rho\BV q)\bigg)=0,
\end{equation}
\begin{multline}
\parderiv{(\rho\BM S)}t+\nabla\cdot(\BV u\otimes\BM S)+\rho\bigg(\BM S
\nabla\BV u+\nabla\BV u^T\BM S-\frac 23(\BM S:\nabla\BV u)\BM I \bigg)
+\frac\Rey{\Mach^2}\bigg[\nabla\cdot(\rho\BM Q)+\\+\frac 25\bigg(
  \nabla(\rho\BV q)+\nabla (\rho\BV q)^T-\frac 23\nabla\cdot(\rho\BV
  q)\BM I\bigg)+\rho\theta \bigg(\nabla\BV u+\nabla\BV u^T-\frac 23
  (\nabla\cdot\BV u)\BM I\bigg)\bigg] =\BM 0,
\end{multline}
\end{subequations}
where $\Mach$ and $\Rey$ are the Mach and Reynolds numbers,
respectively:
\begin{equation}
\Mach = \frac U{\sqrt T},\qquad\Rey=\frac{U L}\Nu.
\end{equation}
Here, for convenience, our definition of the Mach number differs from
the conventional one by the adiabatic constant factor.

As in \citep{Abr21}, here we assume that $\Mach\sim 1$,
$\Rey\to\infty$.  Subsequently, the terms with the shear stress $\BM
S$ are discarded from the momentum and pressure equations (as they are
divided by $\Rey$). The closure for the heat flux $\BV q$ follows from
the stress equation, where we set the terms, multiplied by $\Rey$, to
zero, and take into account the fact that the trace of $\BM Q$ along
any pair of its indices is zero:
\begin{subequations}
\begin{equation}
\frac 25\big(\nabla(\rho\BV q)+\nabla(\rho\BV
q)^T\big)+\nabla\cdot(\rho\BM Q)=-\rho\theta\big(\nabla\BV
u+(\nabla\BV u)^T\big),
\end{equation}
\begin{equation}
\label{eq:heat_flux_closure}
\frac 25\nabla\cdot(\rho\BV q)=-\rho\theta\nabla\cdot\BV u.
\end{equation}
\end{subequations}
Recall that the Newton law of viscosity and the Fourier law of heat
conduction, which are used to close the conventional hierarchy of
velocity moments of the Boltzmann equation, are obtained from the
truncated Hilbert expansion of $f$ in Hermite polynomials around the
equilibrium \citep{ChaCow,Gra,HirCurBir}. To be accurate, this
truncation requires the distribution $f$ to be sufficiently close to
the Maxwell--Boltzmann equilibrium state. In contrast, our high
Reynolds number closure above makes no assumption on the form of $f$
in \eqref{eq:vlasov}, and does not require $f$ to be related to the
equilibrium state in any particular way.

Substituting \eqref{eq:heat_flux_closure} into the pressure equation
\eqref{eq:pressure}, and reverting back to the dimensional variables
by discarding the Mach number, we arrive at the following closed
system of macroscopic turbulent equations for a short range
interaction potential:
\begin{subequations}
\label{eq:turbulent_equations}
\begin{equation}
\parderiv\rho t+\nabla\cdot(\rho\BV u)=0,\qquad\parderiv{(\rho\BV
  u)}t+\nabla\cdot(\rho\BV u^2)+\nabla(\rho\theta+\bar\phi)=\BV 0,
\end{equation}
\begin{equation}
\label{eq:pressure_temperature}
\parderiv{(\rho\theta)}t+\BV u\cdot\nabla(\rho\theta)=0\qquad
\text{or}\qquad\parderiv{(\theta^{-1})}t+\nabla\cdot(\theta^{-1}\BV
u)=0.
\end{equation}
\end{subequations}
In \eqref{eq:pressure_temperature}, either of the two equations can be
used to compute the solution; however, for a numerical simulation, the
inverse temperature equation is more practical as it has the form of a
conservation law. The notable difference between
\eqref{eq:turbulent_equations} and the conventional Euler equations is
that the former have an additional forcing term $\nabla\bar\phi$ in
the momentum equation, and preserve the pressure $\rho\theta$ along
the streamlines.

In \citep{Abr21}, we further imposed the hydrostatic balance
assumption onto the respective analog of
\eqref{eq:turbulent_equations}.  This was necessitated by the fact
that, in the context of \citep{Abr21}, the transport equations were
forced directly by the interaction potential $\phi$, which usually
peaks strongly near zero and thus can force the velocity to attain
unrealistic values. In the present context, however, the momentum
equation in \eqref{eq:turbulent_equations} is forced by the mean field
potential $\bar\phi$, which is given via \eqref{eq:mean_field}, and
behaves regularly throughout the domain. In addition to this, in
practical scenarios, the effect of $\bar\phi$ is expected to be weak,
so one cannot realistically expect a hydrostatic balance to hold for
\eqref{eq:turbulent_equations}. Thus, in the context of the present
work, we abstain from further simplifications of
\eqref{eq:turbulent_equations}.

We also have to note that the equations in
\eqref{eq:turbulent_equations} are ``pathologically turbulent'', in
the sense that, by construction, they cannot model anything but a
turbulent flow with a high Reynolds number. For example, an airfoil,
being moved in an otherwise steady gas at a constant pressure, will be
unable to create, in the context of \eqref{eq:turbulent_equations},
the aerodynamic lift (or the drag, for that matter), due to the fact
that the pressure remains constant along the streamlines. The
conservation of the pressure along the streamlines is the direct
consequence of the high Reynolds number condition requirement, which
is used to achieve the heat flux closure in
\eqref{eq:heat_flux_closure}.

\subsection{Special case: turbulent flow at constant pressure}

A notable special case of the turbulent flow equations in
\eqref{eq:turbulent_equations} is where the pressure is constant
throughout the domain. This occurs for example, in the case of a
simple channel flow at a constant rate (such as the Bernoulli flow),
which also happens to be a steady state for the conventional Euler
equations. For a constant pressure $\rho\theta$, the turbulent flow
equations in \eqref{eq:turbulent_equations} become
\begin{equation}
\label{eq:constant_pressure_equations}
\parderiv\rho t+\nabla\cdot(\rho\BV u)=0,\qquad\parderiv{(\rho\BV
  u)}t+\nabla\cdot(\rho\BV u^2)+\nabla\bar\phi=\BV 0.
\end{equation}
Although the pressure is no longer a variable in the equations, it
still affects the dynamics as a parameter. For example, in the case of
a hard sphere mean field potential \eqref{eq:mean_field_HS}, the
momentum equation is given via
\begin{equation}
\label{eq:constant_pressure_HS}
\parderiv{(\rho\BV u)}t+\nabla\cdot(\rho\BV u^2)+\frac{p_0}{
  2\rho_{sp}}\nabla\rho=\BV 0,
\end{equation}
where $p_0$ is the constant pressure parameter. It is clear that the
magnitude of the constant pressure parameter affects the potential
forcing, and, therefore, the development of the resulting turbulent
effects.

\section{Numerical simulation}
\label{sec:numerics}

Here, we show the results of a numerical simulation of the turbulent
flow of an argon-like hard sphere gas in a straight pipe of square
crossection at normal atmospheric conditions, assuming that the
interaction potential is that of a hard sphere
\eqref{eq:mean_field_HS}. The constant sphere density parameter in
\eqref{eq:mean_field_HS} is set to $\rho_{sp}=2716$ kg/m$^3$, which is
estimated from the mass of an atom of argon ($6.634\cdot 10^{-26}$ kg)
and its van der Waals diameter ($\sim 3.6\cdot 10^{-10}$ meters). The
length of the pipe is 10 meters, while the width and the height are
both set to 2.5 meters.

To carry out the numerical simulations, we use OpenFOAM
\citep{WelTabJasFur}. Since the turbulent flow equations in
\eqref{eq:turbulent_equations} comprise a system of conservation laws,
we implement them with the help of an appropriately modified
\texttt{rhoCentralFoam} solver \citep{GreWelGasRee}, which uses the
central scheme of \citet{KurTad} for the numerical finite volume
discretization, with the flux limiter due to \citet{vanLee}. The
time-stepping of the method is adaptive, based on the 20\% of the
maximal Courant number, which is evaluated over the whole domain at
each time step. The spatial domain is discretized uniformly in all
directions, comprising 400 points along the length of the pipe, and
100 points in each of the two transversal directions (4~million
discretization cubes in total).

We choose the initial condition to be a steady (in the context of the
conventional Euler equations), constant pressure, laminar Bernoulli
jet, with the velocity profile given via
\begin{equation}
\BV u=(u(r),0,0),\qquad u(r)=\left\{\begin{array}{l@{\qquad}l}
u_0(1+\cos(\pi r/r_0)), & r<r_0, \\ \text{zero}, &
\text{otherwise}.\end{array}\right.
\end{equation}
Above, $r$ is the distance to the central axis of the pipe, and the
parameters are chosen to be $u_0=10$ m/s, and $r_0=0.5$ m. As we can
see, the velocity of the jet equals 10~meters per second at the axis
of the pipe, and smoothly decays to zero at the distance of half a
meter away from the axis. Our set-up corresponds to the Reynolds
number in the $\Rey\sim 10^6$ range; indeed, the size of the domain is
$\sim 1$ meter, the reference velocity is $\sim 10$ meters, while the
kinematic viscosity of argon at normal conditions is $\sim
10^{-5}$~m$^2$/s.

The constant pressure is set to $p_0=10^5$ kg/m s$^2$, which
corresponds to the normal atmospheric pressure at sea level. The
temperature $\theta$ is calculated according to Bernoulli's law for
monatomic gases:
\begin{equation}
\theta(r)=\frac{RT_0}M-\frac 15u^2(r).
\end{equation}
Above, $R=8.314$ J/mol K is the universal gas constant, $M=3.99\cdot
10^{-2}$ kg/mol is the molar mass of argon, and $T_0=288.15$ K is the
atmospheric temperature at normal conditions. The density of the
initial flow is calculated according to
\begin{equation}
\rho(r)=\frac{p_0}{\theta(r)}=\frac{5Mp_0}{5RT_0-Mu^2(r)}.
\end{equation}
The boundary conditions are specified as follows:
\begin{itemize}
\item For the velocity $\BV u$, the inlet and wall boundary conditions
  are fixed at the values which correspond to the initial flow, while
  the outlet boundary condition is set to zero normal gradient;
\item For the pressure $p$, the inlet and wall boundary conditions are
  set to zero normal gradient, while the outlet boundary condition is
  set to constant pressure corresponding to the initial flow;
\item For the inverse temperature $\theta^{-1}$, the boundary
  condition is set to zero normal gradient everywhere.
\end{itemize}
Effectively, this set-up corresponds to the constant-pressure
equations in \eqref{eq:constant_pressure_equations}, with the density
boundary condition set to zero normal gradient.

\begin{figure}
\includegraphics[width=\textwidth]{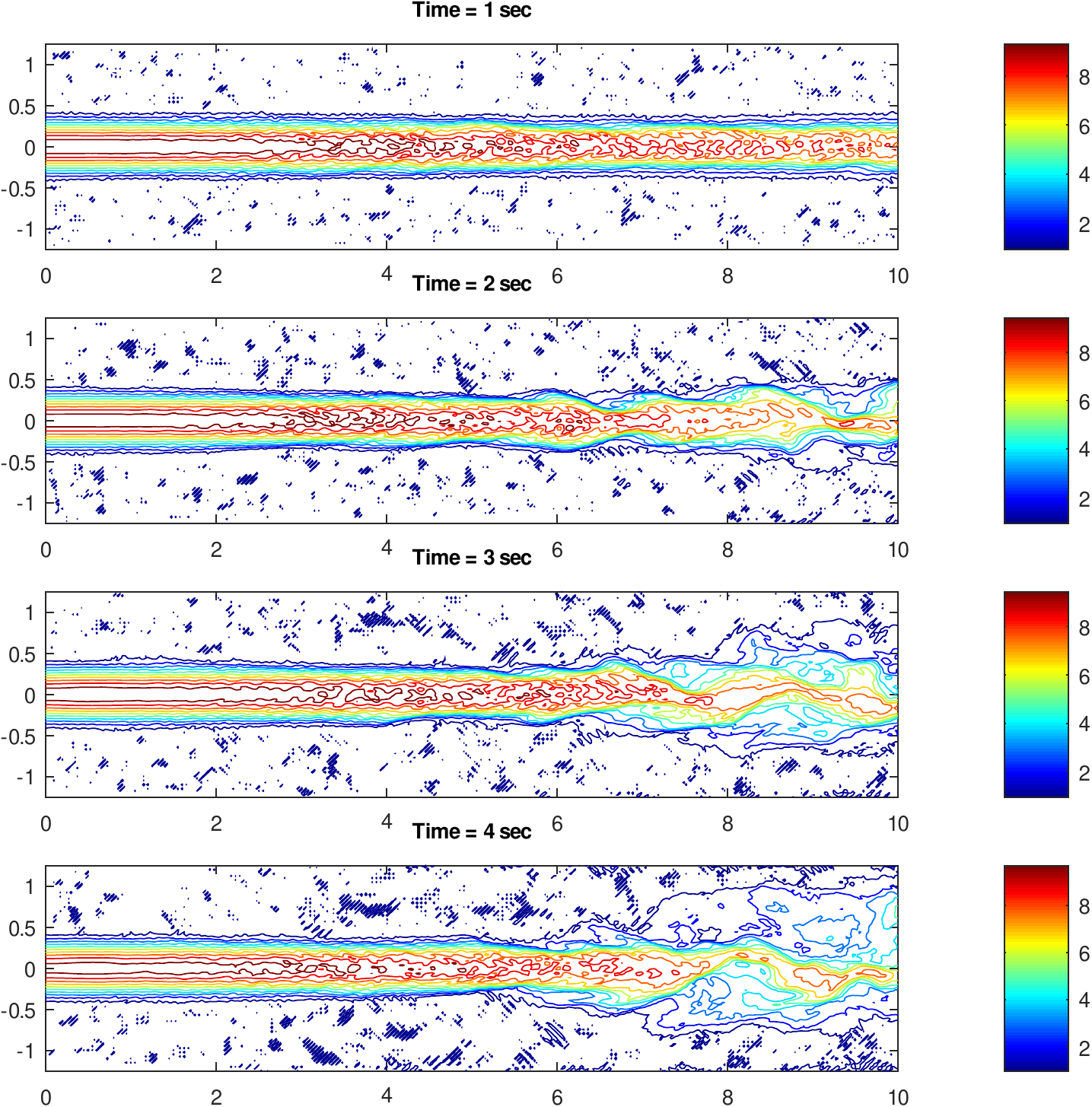}
\caption{The velocity magnitude (m/s) at the elapsed time $t=1,2,3,4$
  sec.}
\label{fig:velocity}
\end{figure}

Upon conducting the numerical simulation, we found that, for the above
specified conditions, the time scale of development of turbulent flow
from the laminar initial condition is only a few seconds. In Figure
\ref{fig:velocity} we show the two-dimensional snapshots of the
magnitude of the flow velocity, measured within the plane $z=0$ (that
is, the plane that dissects the pipe into two symmetric halves along
its axis) at elapsed times of 1, 2, 3 and 4 seconds. As we can see, at
one second, the jet already has small turbulent fluctuations within
it. At two seconds, the last third of the jet develops larger scale
fluctuations, which eventually lead to a break-up. At three and four
seconds, the jet shows three distinct stages: roughly, the first third
of the jet is intact, the second third has small fluctuations, while
in the last third the jet completely falls apart. Even though we have
not observed this turbulent jet approach a steady state, it appears
that its statistical steady state is practically reached at about
three seconds of the elapsed time -- that is, further simulation shows
a similar principal structure of the jet, consisting of the same three
stages.

\begin{figure}
\includegraphics[width=\textwidth]{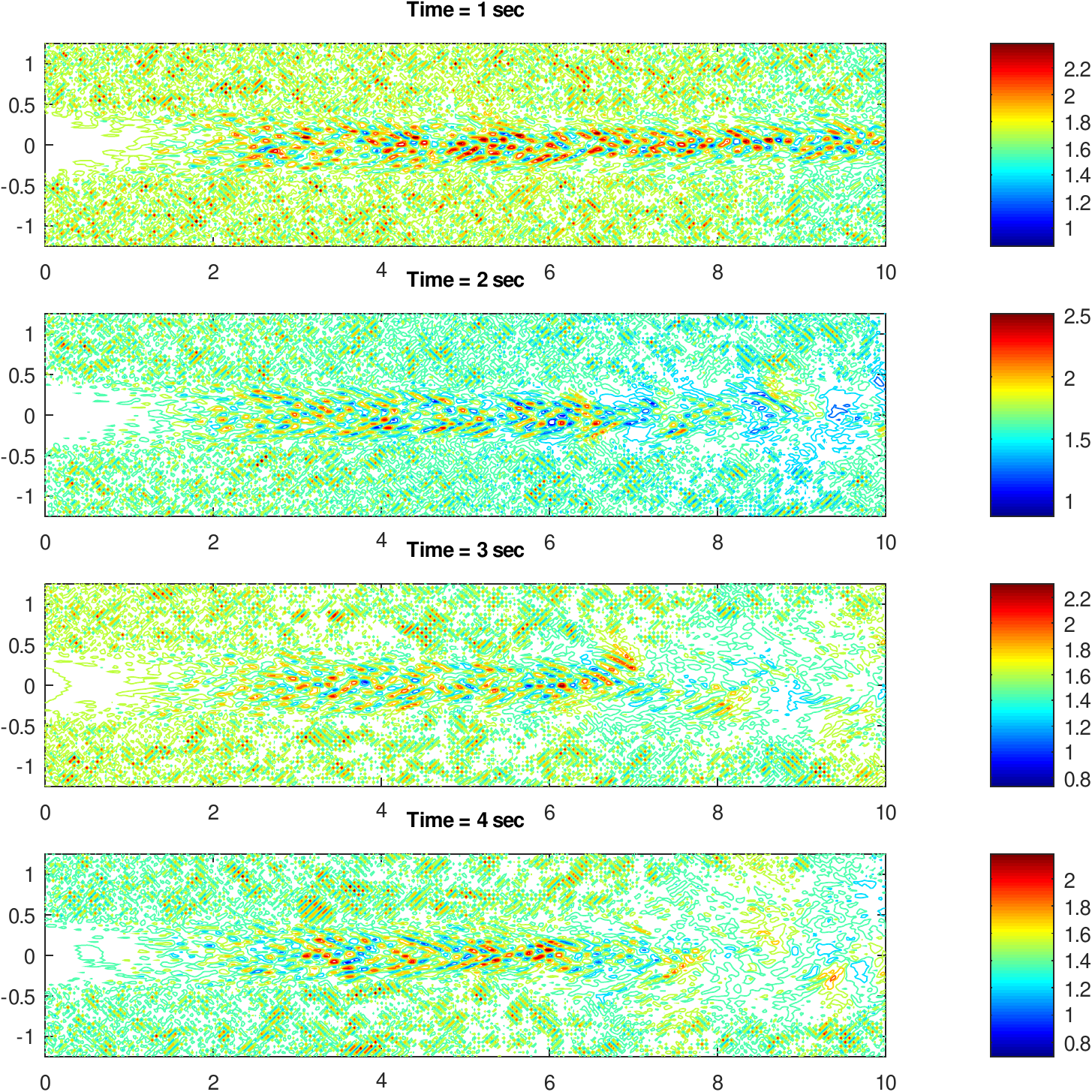}
\caption{The density (kg/m$^3$) at the elapsed time $t=1,2,3,4$ sec.}
\label{fig:density}
\end{figure}

In Figure \ref{fig:density} we show the show the two-dimensional
snapshots of the density, measured within the plane $z=0$ at elapsed
times of 1, 2, 3 and 4 seconds. It is interesting that the most
significant fluctuations of the density occur within the jet itself,
and not in the turbulent flow around it (for example, compare the
snapshots at $t=1$, where the jet is mostly intact throughout the
length of the pipe, and at $t=4$, where the jet falls apart in the
last third of the pipe). The fine ``checkerboard'' fluctuations on the
spatial scale of the discretization above and below the jet appear to
be an analog of the acoustic waves; indeed, combining the
time-derivative of the density equation in
\eqref{eq:constant_pressure_equations} with the divergence of the
momentum equation in \eqref{eq:constant_pressure_HS} leads to
\begin{equation}
\parderiv{^2\rho}{t^2}=\nabla^2:\Big[\Big(\frac{p_0}{2\rho_{sp}}\BM
  I+\BV u^2\Big)\rho\Big],
\end{equation}
where the matrix which multiplies $\rho$ is obviously positive
definite. The ``speed of sound'' here is, however, quite low (in
comparison with the true acoustic waves) -- observe that, for the
given pressure and the density of the atom of argon, we have
$p_0/2\rho_{sp}\sim 20$~m$^2$/s$^2$, plus $\BV u^2$ contributes at
most 100 m$^2$/s$^2$, which yields about 10-11 m/s for the speed of
such a wave in the given conditions. While one can probably devise
numerical methods to suppress these ``acoustic waves'' by filtering
out the appropriate spatial or temporal scales, here we leave them as
they are, since they do not seem to significantly affect the turbulent
dynamics of the jet in Figure \ref{fig:velocity}.

\begin{figure}
\includegraphics[width=\textwidth]{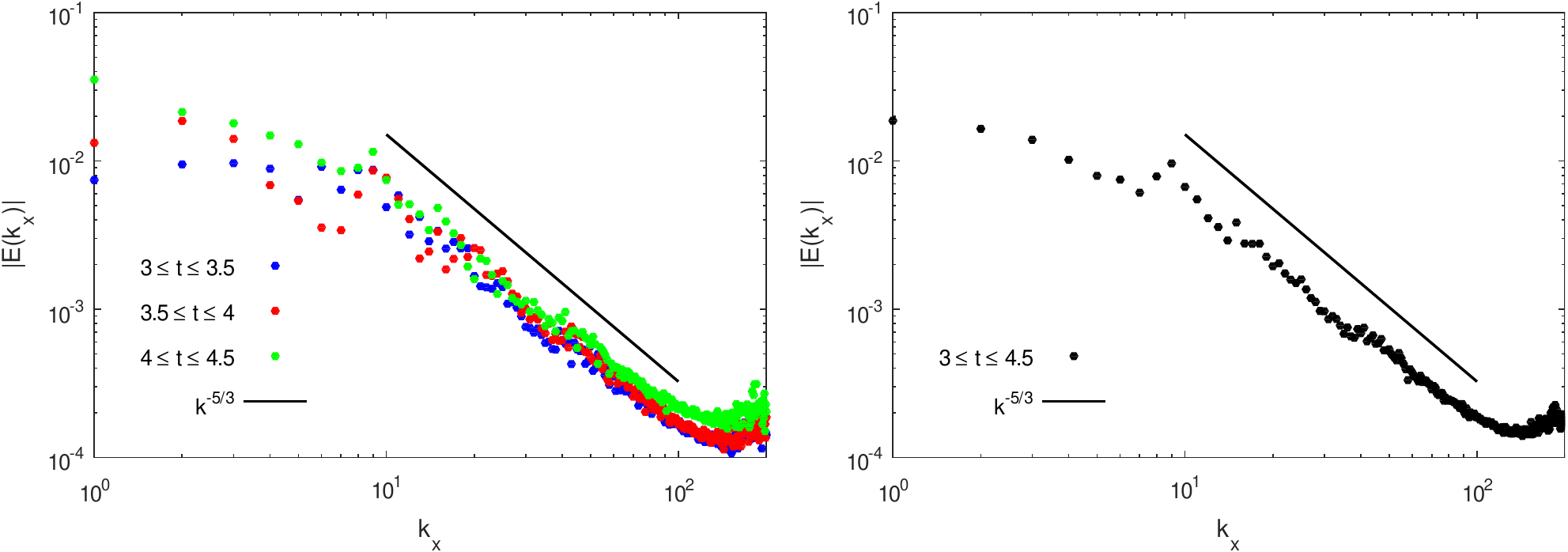}
\caption{The Fourier spectra of the kinetic energy, averaged over
  different time intervals. The slope line $k^{-5/3}$ is given for the
  reference.}
\label{fig:energy}
\end{figure}

To verify that the behavior of the jet in Figure \ref{fig:velocity}
indeed corresponds to the turbulent behavior observed in nature, in
Figure \ref{fig:energy} we show the time-averaged Fourier spectra of
the kinetic energy of the flow. These spectra is computed as follows:
first, the kinetic energy $E(\BV x)=\|\BV u(\BV x)\|^2/2$ is averaged
over the cross-section of the pipe (thus becoming the function of the
$x$-coordinate only). Then, the linear least-squares fit is subtracted
from the result in the same manner as was done by \citet{NasGag}; this
is done to ensure that there is no sharp discontinuity between the
energy values at the inlet and the outlet. Finally, the
one-dimensional discrete Fourier transformation is applied to the
result, so that the resulting Fourier transform of the kinetic energy,
computed along the pipe, corresponds to the zero Fourier wavenumbers
across the pipe. The subsequent time-averaging of the modulus of the
Fourier transform is computed between $t=3$ and $t=4.5$ seconds of the
elapsed time (that is, for the fully turbulent jet). On the left panel
of Figure \ref{fig:energy}, we show the time averages computed for
intervals $3\leq t\leq 3.5$ seconds, $3.4\leq t\leq 4$ seconds, and
$4\leq t\leq 4.5$ seconds. On the right panel of Figure
\ref{fig:energy}, we show the time average for the combined interval
$3\leq t\leq 4.5$ seconds. As we can see, all plots align quite well
with the $k^{-5/3}$ reference line, which indicates the Kolmogorov
decay of the kinetic energy averages. Also, the plots are visually
similar to the observations in the experimental work by
\citet{BucVel}.

\section{Summary}
\label{sec:summary}

In the current work, we develop and test a fluid mechanical model for
a macroscopic turbulent flow, which is triggered by a short range
interaction potential. We start with a system of many identical
particles, which interact via a short range potential. Applying the
BBGKY formalism and a suitable closure, we obtain a Vlasov-type
equation for the distribution density of a single particle with a mean
field forcing. We show that the mean field forcing has an explicit and
relatively simple form if the interaction potential corresponds to a
hard sphere collision.

Next, for this Vlasov-type equation, we formulate the velocity moment
hierarchy, similarly to how it is done in the conventional transition
from kinetic theory to fluid mechanics. However, due to the fact that
the mean field forcing replaces the conventional Boltzmann collision
integral, the closure for the moment hierarchy differs from the one
typically used in the conventional fluid mechanics; namely, we have to
use the high Reynolds number condition for the closure instead of the
Newton and Fourier laws. As a result, we arrive at a system of
transport equations for the density, momentum and temperature, which
is markedly different from the conventional Euler equations. First,
the momentum equation in the new system has an additional forcing
originating from the interaction potential, and, second, the new
system preserves the pressure along the streamlines (while the
conventional Euler equations preserve the entropy).

Finally, we numerically simulate the new equations in a realistic
setting of argon flowing through a straight three-dimensional
pipe. The laminar Bernoulli jet is chosen as an initial condition for
the simulation, given that it is a steady state in the conventional
Euler equations. Our simulation, however, shows that the laminar
Bernoulli jet quickly develops into a fully turbulent flow, which
visually resembles the observations in nature. We also confirm that
the time averages of the kinetic energy of the simulated flow decay
with the rate of negative five-third power of the Fourier wavenumber,
which corresponds to the Kolmogorov spectrum.

\ack This work was supported by the Simons Foundation grant \#636144.

\end{document}